\pdfoutput=1
\documentclass[aps,pre,preprint,groupedaddress]{revtex4-1}

\usepackage{graphicx}
\usepackage{dcolumn}
\usepackage{amsmath}    
\usepackage{amssymb}
\usepackage{bm} 
\usepackage{hyperref}
\usepackage{latexsym}
\usepackage{verbatim}
\usepackage{color}
\usepackage[caption=false]{subfig}

\def\beq{\begin{equation}}
\def\eeq{\end{equation}}

\def\beq{\begin{equation}}                           
\def\eeq{\end{equation}}                           
\def\bea{\begin{eqnarray}}                           
\def\eea{\end{eqnarray}}        

\preprint{}

\begin{document}

\preprint{}

\title{Binary phase separation in a collection of self-propelled particle with variable speed }
\author{Jay Prakash Singh}
\email{jayps.rs.phy16@itbhu.ac.in}
\affiliation{Indian Institute of Technology (BHU) Varanasi, India 221005}
\author{Shradha Mishra}
\email[]{smishra.phy@itbhu.ac.in}
\affiliation{Indian Institute of Technology (BHU) Varanasi, India 221005}
\date{\today}
\begin{abstract}
	{ We study the collective behavior of binary mixture of self-propelled particles. Particles
	moves along their heading direction with {\it variable speed} and interact through short 
	range alignment interaction.
	A variable speed parameter $\gamma >0$ is introduced such that for 
	$\gamma=0.0$ model reduces to {\it constant speed} Vicsek's model. 
	We mix the particles with two different $\gamma$'s  and study the steady state
	behavior of the mixture for different choice of $\gamma$'s and noise strength. One of the 
	$\gamma$ is kept fixed to $1.0$ and another one is varied from small $0.0$ to larger values $8.0$.
	Properties of system is characterise by two types of order parameters (i) orientation order
	parameter, which is a measure of ordering in the system and (ii) density order parameter, which
	measures the phase separation is the system. 
        For all set of $\gamma$'s, 
	system shows a transition from disorder-to-ordered state on the variation of 
	noise strength. The nature of transition and critical noise
	is independent of value of $\gamma$, which is also supported from coarse-grained
	hydrodynamic study. On the variation of system parameters, ($\gamma$'s, $\eta$), we find four distinct phases, 
	(i) ordered phase separated, (ii) ordered mixed, (iii) disordered mixed and (iv)
	disordered phase segragated. 
	Our study shade light on different phases of mixture of different types of active particles.}

\end{abstract}
\maketitle
\section{Introduction \label{introduction}}
\begin{figure}[ht]
\centering
\includegraphics[width=0.8\linewidth]{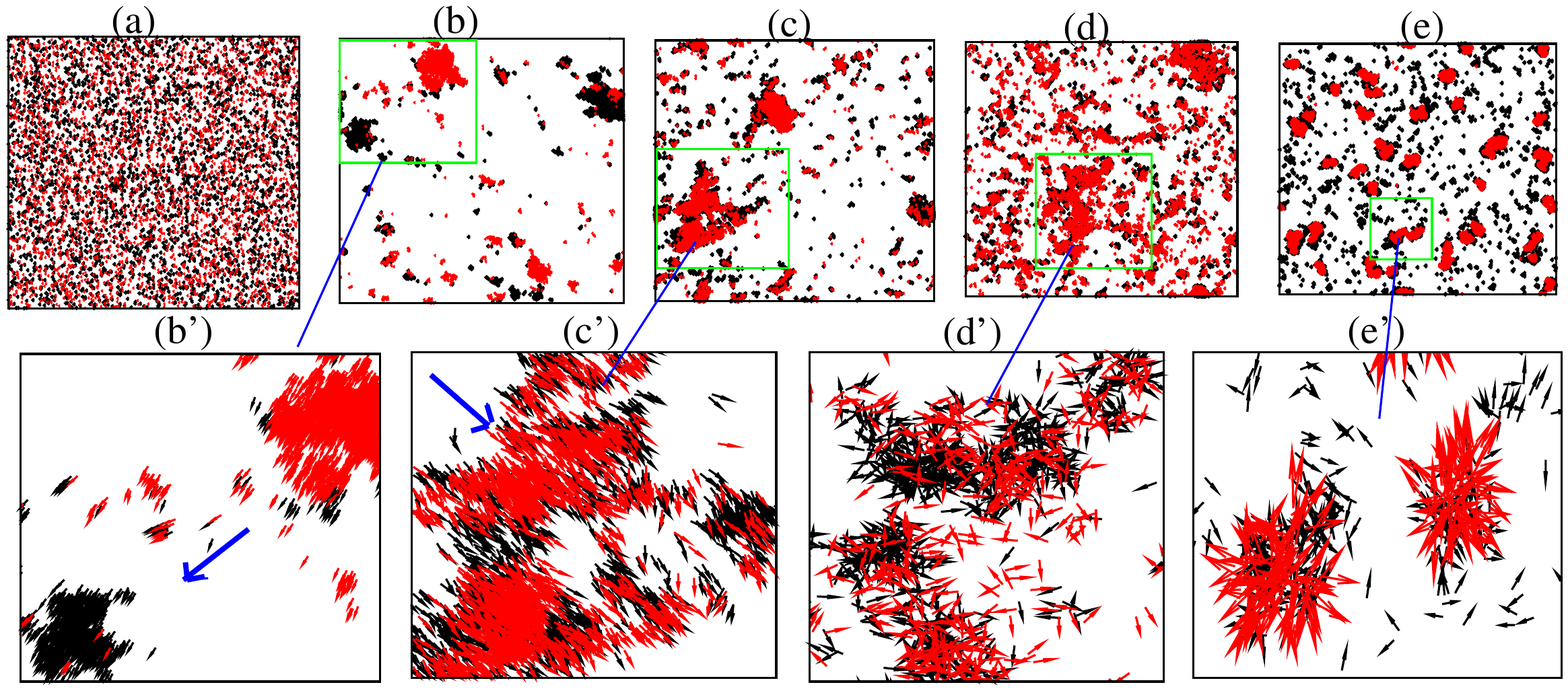}6
	\caption{(color online) Top panel: real space snapshot of particle position of two types of particle with the direction 
	of their velocity vector. Color represents two types particle. Black is for particle of type one and 
	red for second type particle. (a) is for initial random homogeneous mixed state, (b) is for ordered
	phase separated state ($\gamma_2=8$, $\eta=0.2$), (c) ordered mixed ($\gamma_2=0.5$ $\eta=0.2$, (d) is for disordered mixed $(\gamma_2=0.5$, $\eta=0.62$) and (e) is for
	disordered phase segregated phase ($\gamma_2=8$, $\eta=0.62$). 
	Bottom panel (b'-e') are the zoomed version of top panel plot for better
	clarity of four different phases. All snapshots are collected in the steady state and plots are a part of full system.
	Other parameters are same as given in Fig. \ref{fig:fig1}.}
\label{fig:fig1}
\end{figure}

Collection of polar self-propelled particle ubiquitous \cite{physicstoday, fishschool}. 
Examples ranges from very small  intracellular scale to much larger scale \cite{sriramrev3, sriramrev2, sriramrev1, harada, badoual, nedelec, rauch, ben, appleby, helbing, helbing1, physicstoday, kuusela, hubbard, schaller, sumino, peruani, bacterialcolonies}. 
Study of such system started with the novel work of T. Vicsek \cite{vicsek}, In this study,
each individuals are modeled as point particle move along their heading direction  with a
{\it constant} speed and align through a short range alignment interaction with their neighbors. Interestingly 
different variants of Vicsek's model is studied but mainly with constant speed 
\cite{chatepre2008, chate2007, katz, shradhasudipta, shradhamanna}. 
But in reality there is no reason for the speed of particles to be fixed. For examples 
in everyday traffic, car can not move if stuck in jam situation but move freely,
when other vehicles are moving in the same direction.  Not only in everyday traffic but 
experiments on living bacteria
 {\it Bacillus Subtilis} observed that speed of each individual depends on
polarisation of their neighbors \cite{goldstein2012pre}. 
Our previous study is motivated by an  experiment on fish school:  and  a variable speed model
is introduced in \cite{shradhapre2012}.  A variable speed model is introduced, where 
speed of the particle depends on their local neighbors 
orientation through  a variable speed parameter $\gamma >0$ (with a {\it power-law}). For any 
$\gamma>0$ when particle moves in well ordered
region then its speed is maximum and in the disordered region speed is close to zero. For $\gamma=0$,
all the particle moves with {\it constant} speed. Hence model
is very much applicable for  situations where random moving  crowd restrict the motion of particle. The variable
speed parameter $\gamma$ introduced here can be thought of as characteristics of particles, which gives how 
particle response to its neighbors. It can have origin from various biological or physical factors.
In this article we will not go into details of such factors. We will strictly consider a variable
speed model introduced in \cite{shradhapre2012}. And  ask the question what 
happens if we mix the particles with two different values of variable
speed parameters ($\gamma_1, \gamma_2$)? Whether we find  a phase separation  
for certain range of system parameters, {\it viz.} noise 
strength and $\gamma$.\\ 
In this study one of  the $\gamma_1=1$ is fixed, {\em i.e.} speed of the 
particle linearly vary with local neighbor's alignment.  And other $\gamma_2$ is tuned from $0$ to $8$. 
Experiment
on fish-school (Golden-shiner) found that speed of the fish depends on local
neighbor's alignment with  variable speed parameter $\gamma=6$ \cite{shradhapre2012}. 
Hence we expect for other type particle one  will have different $\gamma$. 
Properties of the system is characterise by two types of order parameter. (a)  orientation
order parameter (OOP)  $\chi$, which is a measure of global orientation of the flock and 
(b) density order parameter (DOP) $\phi$, 
which is measure of phase separation among two-types of particles.
We first measure the $\chi$ as a function of noise strength for different values of
variable speed parameter $\gamma_2$. For  set of $\gamma$'s = ($\gamma_1=1, \gamma_2=0-8)$ we find a transition 
from disordered-to-ordered state on the variation of noise strength, critical noise (is close to $0.6$)
is almost independent of  
 the variable speed parameter $\gamma$. Which is further confirmed by the 
mean-field analysis of the coarse-grained hydrodynamic equations of motion for slow 
variables.
On the variation of two parameter ($\gamma_1, \gamma_2)$ and noise strength we find four distinct phases.
(i) For small noise when system is globally ordered $\chi \simeq 1$ and  $\gamma_2 > 3$:
the particles with two different $\gamma$'s  are phase separated   and $\phi >0.6$. Hence they move in 
the group of their own
types of particles. Typical snapshot for small $\eta=0.2$ and  $\gamma_2=8$ is shown in 
Fig. \ref{fig:fig1} (b) and (b').  We call this phase as ordered-phase separated phase (OPS). 
(ii) Again for small noise when $\chi$ is large but  
$\gamma_2$ is close to $\gamma_1=1$, phase separation decreases $\phi <0.6$. This is 
defined as ordered mixed phase (OM). Please see the snapshot Fig. \ref{fig:fig1}(c) and (c').
As we increase noise strength  and cross the  ordered region $\eta>0.6$, 
we again find two different phase (iii) disorder mixed (DM) and disordered phase segregated (DPS) 
when difference is two $\gamma$'s is smaller/larger that $2$. 
Please see the snapshot shown in Fig. \ref{fig:fig1}(d-e) and (d'-e').
In Fig. \ref{fig:fig3} we plot the DOP {\it vs.} $\gamma_2$ for two different noise strengths
(a) $\eta=0.2$ in the ordered
region and (b) $\eta=0.62$ in the disordered region. 
We draw the four  phases with different shaded regions. Which shows the value of
DOP for four distinct phase we find here.
In rest of the article we discuss the four phases in detail and
also compare the result with hydrodynamic equations of motion.\\
Rest of the article is divided in following manner. In section \ref{model} we discuss our model and 
numerical details of the simulation. Section \ref{results} contains the 
 result of numerical study and in section \ref{Mean-field order-disorder transition using coarse-grained hydrodynamics} and \ref{Linearised study of hydrodynamic equations of motion} we compare  the result with coarse-grained
hydrodynamic equations of motion and finally  section \ref{Discussion} concludes the results and shows final outcome
of our study.

\section{Model \label{model}}
In our model, system consist of symmetric binary mixture  of $N-$ point particles moving on a two-dimensional
substrate. 
Each particle is defined by its position $ r_i(t)$, velocity vector $v_i(t)$.
The velocity of the particle is defined by its unit direction or orientation 
${\bf n}_i(t) = (\cos(\theta_i(t)), \sin(\theta_i(t))$ and  speed $v_i(t)$.
The particles interact through a short range alignment interaction. 
Self-propulsion is introduced as a motion towards its orientation with a variable speed ($v_i(t)$ in unit time).
Unlike the previous models \cite{vicsek, chatepre2008}, here the speed of the particle depends 
on its neighbors. Hence a variable speed model is introduced \cite{shradhapre2012}. 
We first update the position of the particle
\begin{equation}
r_i(t + 1) = r_i(t) + v_i(t){\bf n_{i}}(t)    
\label{eqn1}
\end{equation}
and the orientation update equation with a  short range alignment interaction
\begin{equation}
{\bf n_{i}} (t+1)=\frac{\sum _{j\in R_{0}} {\bf n_{j}}(t)+N_{i}(t){\bf \eta}_{i}}{W_{i}(t)}
\label{eqn2}
\end{equation} 
where in the nominator sum is over all the particles within the interaction radius $R_0 $ of the $i^{th}$ particle, i.e.,
$\vert {\bf r}_{j}(t)-{\bf r}_{i}(t)\vert < R_0$ . $N_i(t)$ is the number of particles within the interaction 
radius of the  $i^{th}$ particle
at time $t$. $W_i(t)$ is the normalisation factor, which  make the R. H. S. of Eq. \ref{eqn2} again a unit vector.
The strength of the noise $\eta$ is varied between
$0$ to $1$. for $v_i(t)=v_0$ model is similar to Vicsek model \cite{vicsek}.
But here unlike the Vicsek's model: we introduce the variable speed: 
guided by the experiments on fish-school, in \cite{shradhapre2012} a variable speed model 
is introduced by considering a
simple power-law relationship between the local polarisation $\chi_i(t)$  around $i^{th}$ particle with
speed $v_i(t)$ such that.
\begin{equation}
v_i(t) = v_{(\chi_i(t))^\gamma}
\label{eqn3}
\end{equation}
where 
\begin{equation}
{\chi_{i}} (t)=\vert\frac{\sum _{j\in R_{0}} {\bf n_{j}}(t)}{N_{i}(t)}\vert
\label{eqn4}
\end{equation}
and $\gamma$, is  variable speed parameter  such that particle moves with maximum speed $v_0$ in well
ordered region and almost static (zero speed) in completely disordered region. For $\gamma=0$, model
reduces to constant speed.
Note that for any 
$\gamma$ an isolated particle will move with
maximal speed $v_0$. 
Hence, the variable speed parameter  $\gamma$ controls the shape of curve that relates local order
and speed. For $\gamma=1$, local speed vary linearly with local polarisation.\\
Here  we consider  a binary mixture of  particles by introducing two parameters ($\gamma_1$ and $\gamma_2$ ) 
of speed such that
$v^i_1(t) = v_0{(\chi(t))^{\gamma_1}}$ and 
$v^i_2(t) = v_0{(\chi(t))^{\gamma_2}}$.
 One of the  $\gamma$,  $\gamma_1$ is fixed to $1.0$ and these particles are called as type one
 and other  $\gamma_{2} $ is varied from $(0,8)$ and particles are called type two.
Agent based numerical simulation is performed with  $N_1$ particles of type one  
and $N_2$ particles of type two ($N_1=N_2=N/2$). 
Started with random mixed state of both types 
particles, all the particles are sequentially updated using the above Eqs. \ref{eqn1},\ref{eqn2},\ref{eqn3}  And it is
counted as one simulation step. 
Simulations  are performed for $10^7$ simulation steps 
with $L=100$  for different values of $\gamma_2= (0,8)$ and noise strength $\eta$. Density
of particle is fixed to $\rho=\frac{N}{L^2}$ and maximum speed
of the particle $v_0=0.5$. For better quality of five different initial realisations are used.\\ 
We study  the system 
for different set of $(\gamma_2, \eta)$. Steady state is characterised by two types
of order parameters: (i) orientation order parameter 
$\chi(t)=\vert\frac{1}{N} \sum _{i=1} ^N {\bf{n}}_{i}(t)\vert.$, which is measure
of orientation of all the particles. When $\chi(t) \simeq 1$ means the ordered state such that large number of particle moving 
in the same direction showing the collective motion. If $\chi(t)=0$ i.e. all the
particle moving randomally in random direction (Disorder).\\ and density 
order parameter (DOP) $\phi=\frac{\sum _{i=1}^N\vert{\rho^i_1(t)}-{\rho^i_2(t)\vert}}{{\sum _{i=1}^N\vert{\rho^i_1(t)}+{\rho^i_2(t)\vert}}}$, which is a measure of phase separation among two types of
particles, where $\rho^i_{k=1,2}(t)$  are the number of particle of type $k$
within the coarse-grained radius of $i^{th}$ particle of same type.
The value of $\phi$ also lies between $0$  and $1$.
when $\phi$ close to $1$, implies only same kind 
of particle inside the interaction radius. Which is possible 
when particles are phase separated.
When $\phi$ is small, hence both types of particles present inside the interaction radius hence mixing.
\begin{figure}[ht]
\centering
\includegraphics[width=0.8\linewidth]{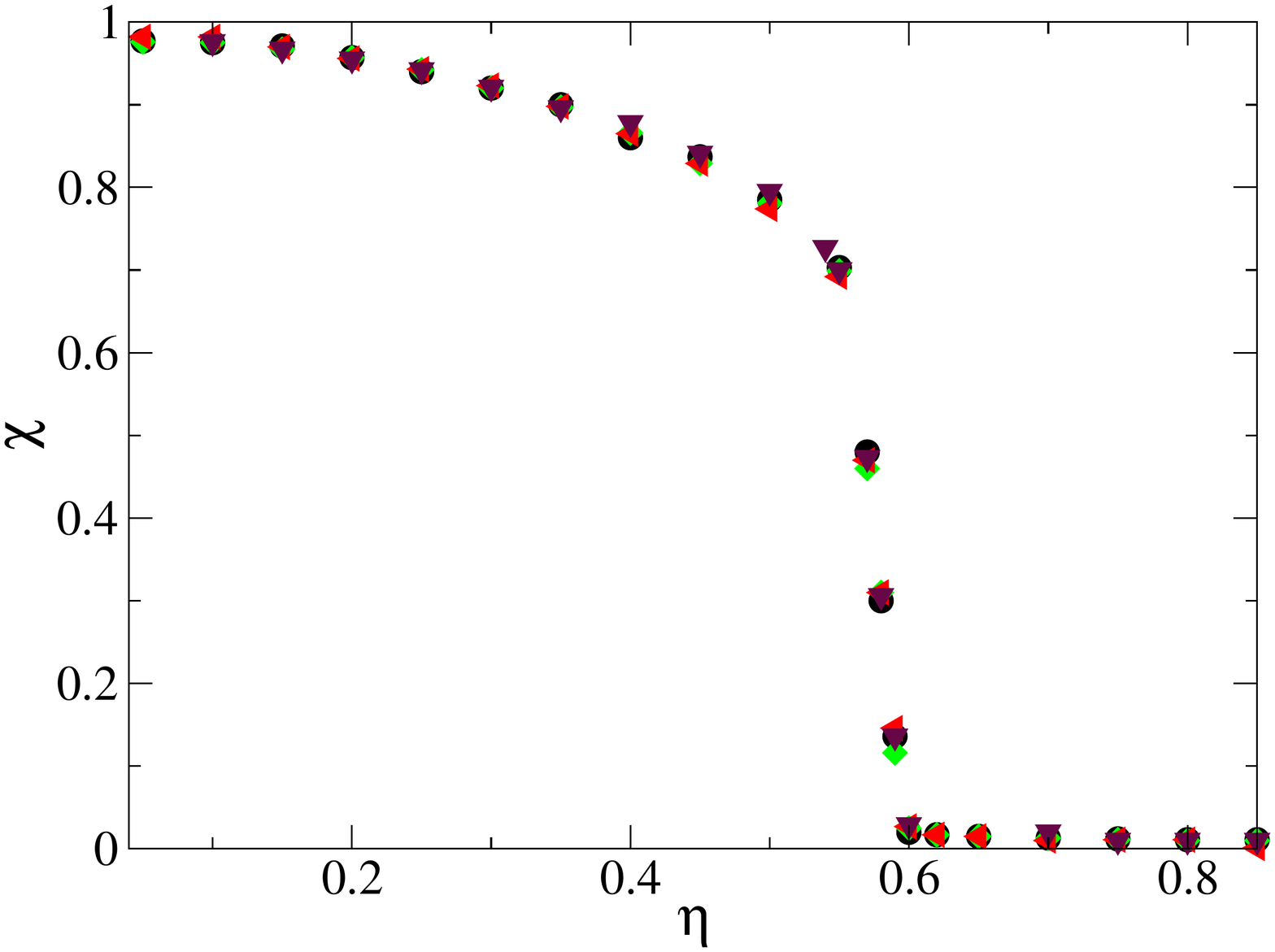}6
\caption{(color online) Plot  of orientation order parameter (OOP)  vs.  noise strength $\eta$,  $\gamma_1=1$, different curves are for
	different $\gamma_2$. $\circ$,  $\square$, $\diamond$ and $\bigtriangleup$ are for   $\gamma_2=2, 4, 6$ and 0.5 respectively. All the curves are independent of $\gamma_2$ and shows a transition from ordered state for small $\eta$ to
	disorder state for large $\eta$. Critical noise lies between $0.57-0.60$. Data is obtained for system size $L=100$ and simulation time $10^7$ in the steady state. Averaging is done over five independent realisations.}
\label{fig: fig2}
\end{figure} 
\section{Results\label{results}}
We first calculate the mean value of $OOP$, $\chi$, averaged over time in 
the steady state and over many realisations. In Fig. \ref{fig: fig2}
we plot the steady state $\chi$ vs. $\eta$ for different $\gamma_2$. 
For all set of $\gamma_2$ we find a transition from disordered random state to ordered state
when $\eta$ is tuned from large  to smaller values. 
For all set of $\gamma_2$
transition remains the same. Hence disorder-to-order transition is independent of variable 
speed parameter $\gamma$. Which is further given in section \ref{Mean-field order-disorder transition using coarse-grained hydrodynamics} using 
the coarse-grained hydrodynamic equations of motion for slow variables. 
We also calculate {\it mean} value of $DOP$ $\phi$, where definition of ``mean'' is same as defined before. 
When the $DOP \simeq 1$, then two species are phase separated from each other and when DOP is 
small then they are mixed. Now we find four types of phases in terms of the two order 
parameters ($\chi, \phi$),
(a) ordered phase separated (OPS), (b) ordered mixed (OM),  
(c) disordered mixed (DM) and (d) disordered phase segregated (DPS).
In Fig. \ref{fig:fig1} we plot the four snapshots for four combination of
($\eta, \gamma_2$). Since for all $ \gamma_2$ disorder-to-order
transition happens at same $\eta$. Hence all of our later measurements are strictly
restricted to ordered $\eta<0.4$ and disordered state $\eta>0.6$. Properties near 
to the disorder-to-order transition is also interesting but it is not of our interest in this work.
 For small $\eta=0.2$ and larger $ \gamma_2>3 $, we find 
$OOP$, $\chi \simeq 1$ and also $\phi\simeq 1$, hence
in the steady state particles  form ordered clusters  and  also phase separated. 
Typical snapshot for this kind of phase is shown in Fig. \ref{fig:fig1}(b) and (b'). 
We name it as order phase separated phase (OPS). 
As we decreases the $\gamma_2$ then the difference in the speed of two types of particle decreases and
they start to mix. Fig. \ref{fig:fig1}(c) and (c') shows one of the typical snapshot of such phase. 
We call such phase as order mixed phase (OM). In this phase $\chi$ is still close to $1$  but $\phi<0.6$. 
Now as we go to the disordered state  $\eta=0.65$ and vary $\gamma_2$. For small
$\gamma_2 <3$, the two types of particles are always mixed and we find no phase separation.
Both $\phi$ and $\chi$ is small, We call this
phase as disorder mixed (DM) and for large $ \gamma_2 > 3$, we find disorder-phase segregated phase (DPS).
The two order parameters $OOP$ and $DOP$ behave similarly for the above two phases but they differ in detail. 
Which we will explain in following subsections.
Typical snapshot of the two phases are shown in Fig.\ref{fig:fig1} (d-e) and (d'-e') respectively. \\
Now we will briefly explain characteristic of all four phases in detail.
\begin{figure}[ht]
\centering
\includegraphics[width=0.8\linewidth]{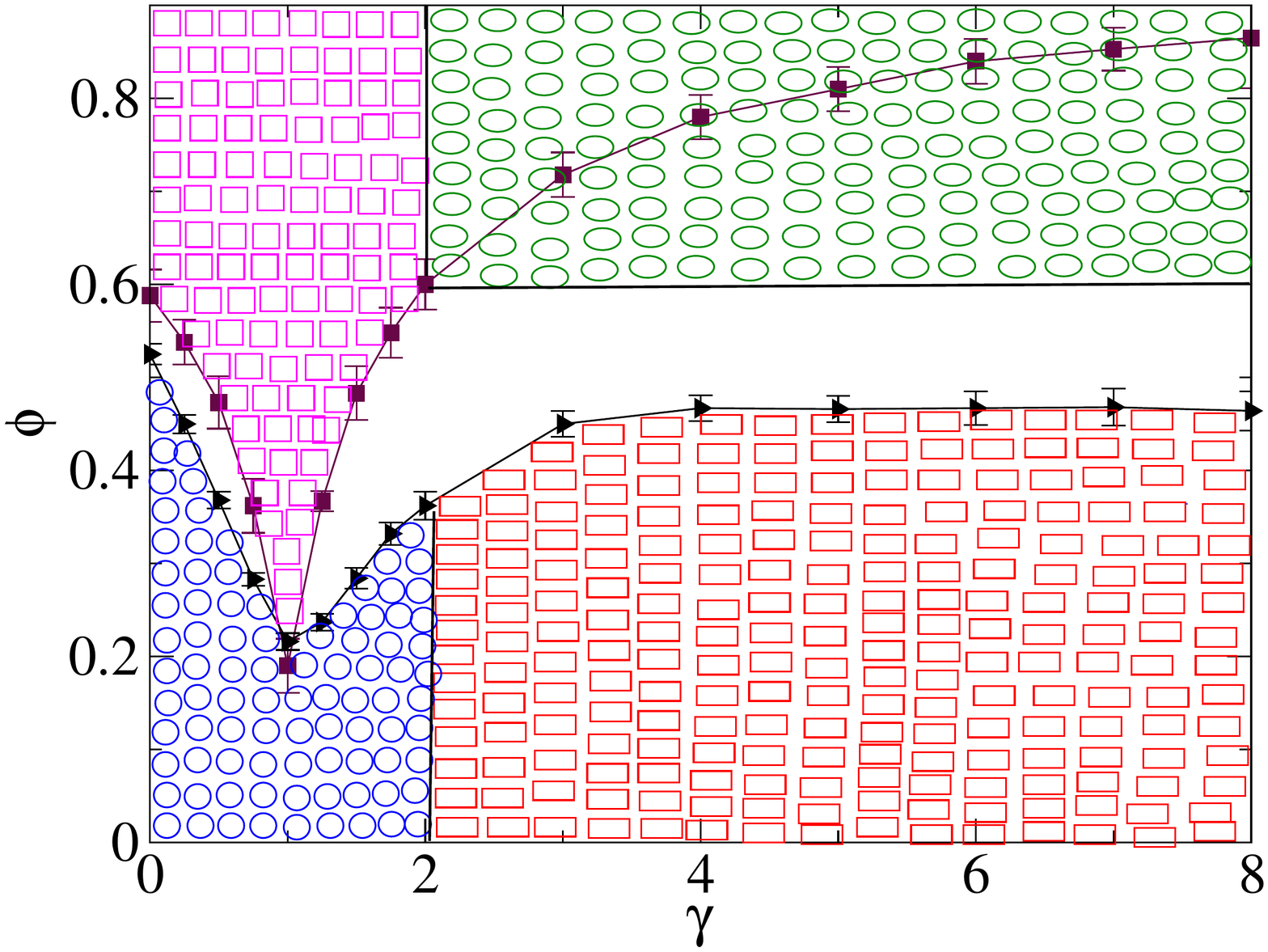}6
	\caption{(color online): Plot of DOP vs. $\gamma_2$ for two different $\eta=0.2$ ($\circ$) and $0.62$ ($\square$). Other parameters
	are same as defined in Fig. \ref{fig:fig1}. 
	Different shaded regions represents the four different phases in the system. 
	(i) Green ellipses represents OPS for ($2<\gamma_2$, $\eta<0.5$). (ii) Magenta square shows OM phase ($0<\gamma_2<2$, $\eta<0.5$). 
	(iii) Blue circles region represents DM phase for $\eta>0.6$ and $0<\gamma_2<2$ and  (iv) Red rectangles for DPS phase for ($2<\gamma_2$, $\eta>0.6$). Empty regions are near to the disorder-to-order phase and phase separation transition, which is not explored in detail in current work.}
\label{fig:fig3}
\end{figure}
\subsection{Ordered phase separated: OPS}

\begin{figure}[ht]
\centering
\includegraphics[width=0.8\linewidth]{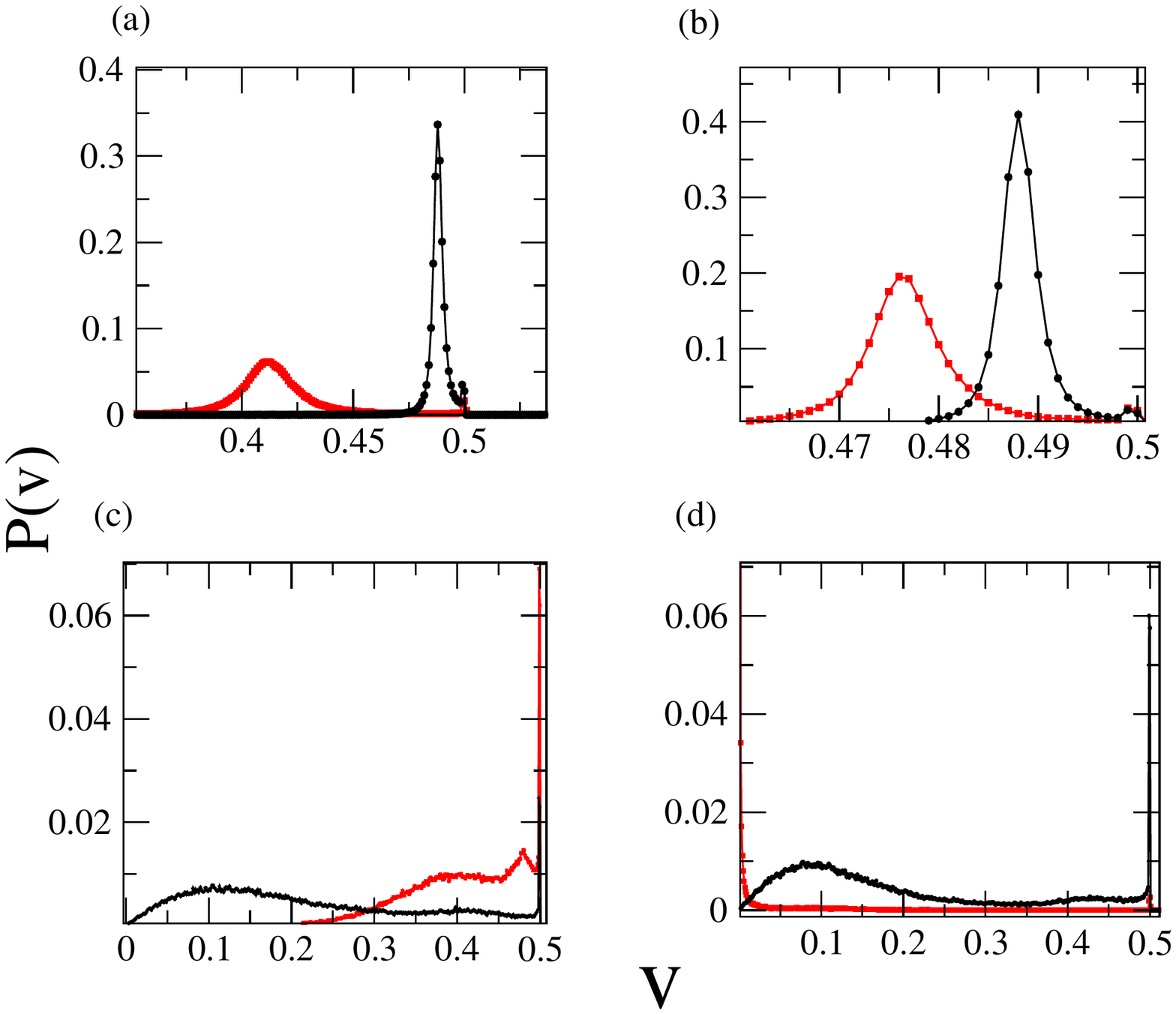}
\caption{Plot of PDF of speed $P(v)$ vs. $v$ for four different phases (a-d) for OPS, OM, DM and DPS respectively. The two 
curves are for $P(v)$ for two types of particles. Symbols have same meaning as in Fig. \ref{fig:fig7}. For all plots 
there is always a peak at maximum speed $v=v_0=0.5$ and second peak is at smaller value. The difference in two peak 
position is large for large $\gamma_2$ in ordered state.  
Other parameters are same as in Fig.\ref{fig: fig2} }
\label{fig:fig4}
\end{figure}

For  $\eta <0.4$ and large $\gamma_2$, the two order parameters 
$OOP$ and $DOP$ are close to $1$. In Fig.\ref{fig:fig3} we plot the
$DOP$ vs. $\gamma_2$ for two different $\eta=(0.2,0.62)$ values. 
For small $\eta=0.2$, starting from initially random and mixed state in the steady state, 
both types of particles
forms moving clusters but they  move in different clusters.
For larger $\gamma_2$ and smaller $\eta$, clusters are more separated and
as we increase $\eta$ and decrease $\gamma_2$ 
phase separation decreases as shown in Fig. \ref{fig:fig3}.
To further understand such phase separation we calculate the probability distribution function (PDF) of
particle speed $P(v)$ for two types of
particles. In Fig.\ref{fig:fig4}(a)
we plot the $P(v)$ for both types of particle for different values of $\gamma_2=8$ and 
$\gamma_1=1$ and for noise strength $\eta=0.2$. $P(v)$ for
both types of particle show one small peak at maximum possible speed $v=0.5$, which is mainly
due to random moving particles. Another peak is present  at smaller 
speed value $v<0.5$. This is contribution from clusters and it fits well with 
normal distribution (lines are fit to the Gaussian distribution). 
We find  that 
the difference in the two peak position $\Delta v/v_0$ increases as we increase $ \gamma_2$. 
Peak position represent the mean speed of particles inside the cluster. 
In Fig. \ref{fig:fig6}  we plot $\Delta v/v_0$ vs. $\gamma_2$ for $\eta=0.2$.\\
In section \ref{Linearised study of hydrodynamic equations of motion} we show the linearised study of  coarse-grained hydrodynamic 
equations of motion for  density  of two types of particles and polarisation ordered parameter. 
The equations are studied for small fluctuations about homogeneous ordered state for different
value of $\gamma_2$. We find that homogeneous ordered state is unstable for large $\gamma_2$, which 
further supports our numerical result. Which shows the presence of OPS state   
for large $\gamma_2$ and small $\eta$.\\
To further characterise different phases we calculate  
$P_{ij}(n)$, where $(i,j=1,2)$ and $n$ is number of particle.
 Since the two types of particles are phase separated in OPS,  hence the 
 two distributions $P_{12}(n)$ and $P_{21}(n)$ 
 looks similar, and should show sharp decay for large $n$, which is due to 
 less mixing of two types of particles.  Other two distributions
$P_{11}(n)$ and $P_{22}(n)$ have broad distribution which confirms the clustering of same 
types of particles. Please see the Fig. \ref{fig:fig5}(a).
In the lower panel of Fig. \ref{fig:fig5}(a) we plot the $P_{ij}(n)$ on 
$\log-\log$ scale. Which shows that the tail of $P_{12}(n)$ $P_{21}(n)$  fits well with 
power $n^{-\alpha}$ with exponent $\alpha \simeq 2$, but $P_{11}(n)$ and $P_{22}(n)$ 
are better fitted with  $\exp^{(\frac{-n}{n_0})}$ and $n_0 \simeq 40$. 

\subsection{ordered mixed: OM}
\begin{figure}[ht]
\centering
\includegraphics[width=0.8\linewidth]{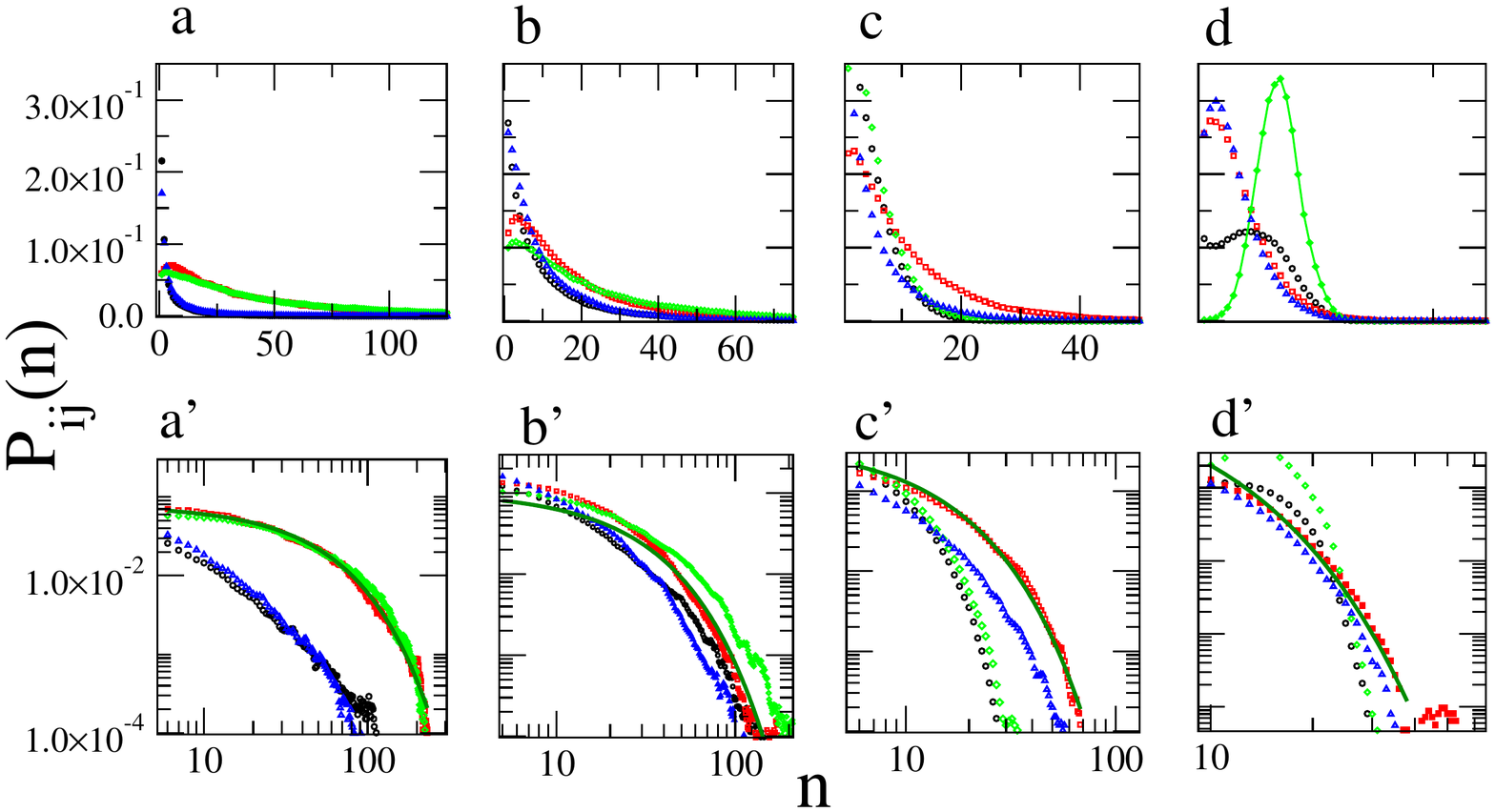}
	\caption{(color online): Plot of particle number PDF $P_{ij}(n)$  for four different phases OPS, OM, DM and DPS,  (a-d) respectively. Top panel is plot on normal scale and bottom is on $\log$-$\log$ scale.  The four curves in each panel is for 
	four distributions as defined in main text. $\circ$, $\square$, $\diamond$ and $\triangle$'s are for $P_{12}(n)$,
	$P_{11}(n)$, $P_{22}(n)$ and $P_{21}(n)$ respectively. In the bottom panel curves are fitted with power-law and exponential
	tail for large $n$.}
\label{fig:fig5}
\end{figure}

\begin{figure}[ht]
\centering
\includegraphics[width=0.8\linewidth]{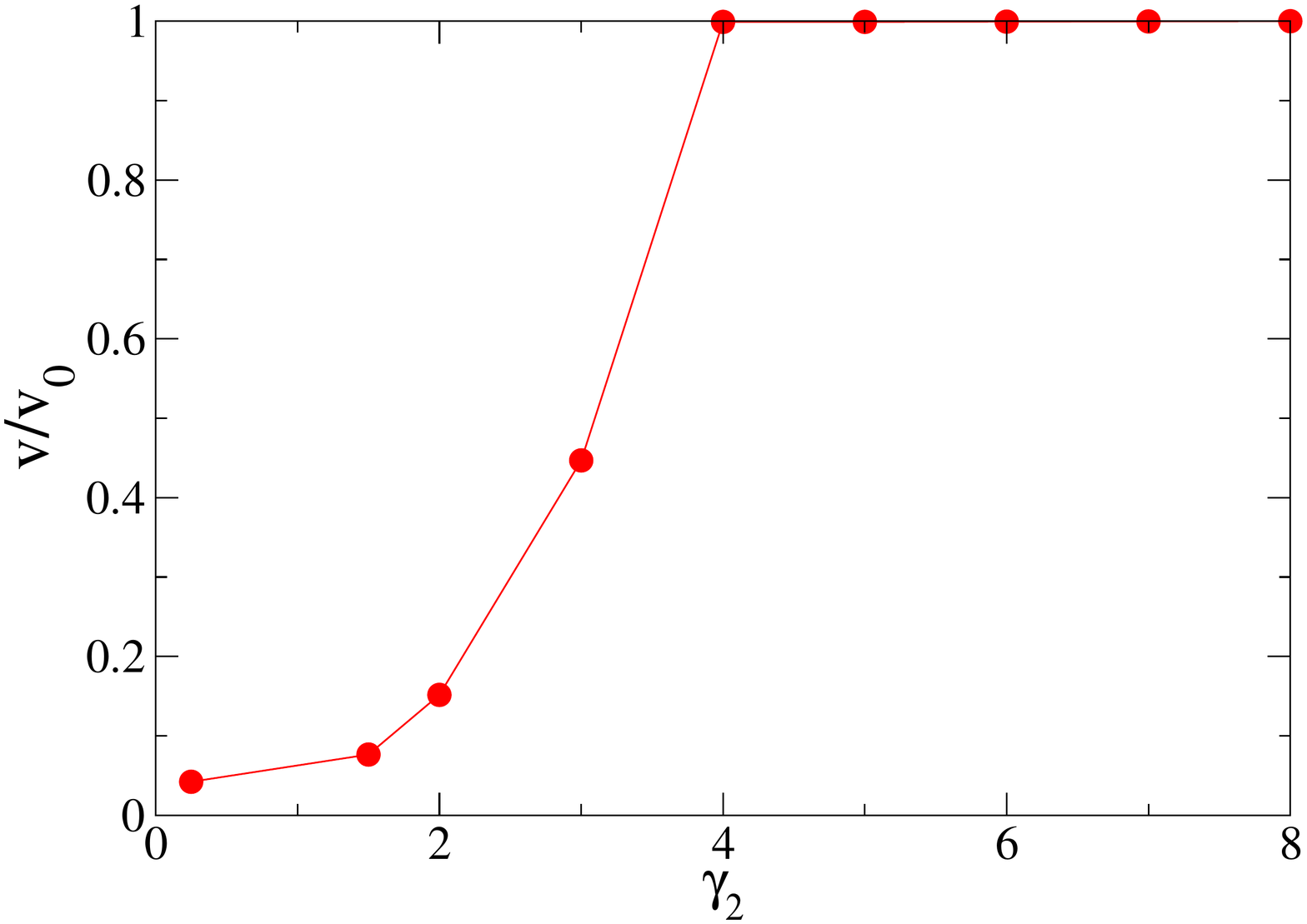}
\caption{(color online): Plot of  normalised velocity difference  $\Delta v/v_0$ vs. $\gamma_2$  for ordered state when $\eta=0.2$.}
\label{fig:fig6}
\end{figure}

In this phase as defined before orientation of particles are aligned along some mean direction
hence $OOP$ is close to $1$, but both types of particles 
remain mix and a cluster consist of both types of particles as shown in Fig. \ref{fig:fig1}(c) and (c'), hence $DOP<0.6$. 
In Fig. \ref{fig:fig4}(b)  we plot
the $P(v)$ for $\gamma_2=0.5$ and  $\eta=0.2$.
We notice two features in the $P(v)$, one peak at $v_0=0.5$ which is again due
to the random isolated moving particles. Second peak appears at $v_0 <0.5$ for
both types of particle. In comparison to previous case when difference in 
two $\gamma$'s $ \gamma_2-\gamma_1$ is large, now difference in the two peak position
decreases also the two distributions starts to overlap as shown in Fig. \ref{fig:fig4}(b).  
The overlap between 
the distributions due to large number of particles of both types moving with same speed. 
Hence they belongs to the same cluster. Again we plot the 
the four $P_{ij}(n)$ in Fig. \ref{fig:fig5}, all four $P_{ij}(n)$  decay exponentially with 
$n_0 \approx 20-30$, which 
implies formation of large clusters.
All four distributions are similar, hence
one type of particle can be in the neighborhood of other type and also of the same type
with equal probability. Which again confirms the mixed phase.

\subsection{Disordered mixed: DM} 

Now we come to the case when 
noise strength is large such that mean orientation of particle is random but 
difference in two types of $\gamma$ is small $\gamma_1=1$ and 
$\gamma_2<3$. In this case both order parameters remain small. Hence
we name the phase as disordered mixed phase. The four number 
distributions are exponential with $n_0 \simeq 3$ for $P_{11}(n)$ and $n_0 \simeq 8$ for 
$P_{12}(n)$ and approximately close to $10$ for $P_{22}(n)$ and $P_{21}(n)$. Size of 
clusters are small in this phase and particles of type one form even smaller clusters.
 $P(v)$ shows broad distribution for both types of
particles and there is very clear overlap. Which further confirms 
the mixing. 

\subsection{Disordered phase segregated: DPS}

Now we tune noise to larger values $\eta>0.6$ and vary the variable
speed parameter  $\gamma_2$. For large $\gamma_2 >3$, i.e.
for  type one $\gamma_1=1$ particle speed vary linearly with local polarisation and for second
type $\gamma_2>>\gamma_1$, 
speed is close to maximum speed for well ordered regions and very small for 
disordered region. In the disordered region when $\eta>0.6$, most of the time 
particles are in disordered cluster region or moving individually. For type one 
particle since
speed vary linearly with local polarisation hence we find a broad distribution of $P(v)$ and another 
type particle, speed can mainly take two possible 
values $0$ and $v_0$ when particle moves individually or in cluster respectively. In Fig. \ref{fig:fig1}(e) and (e')
we plot the real space snapshot of  particle position for both types of particles.
We find that one type of particles, for which $\gamma$ is large forms more or less
static clusters shown in red and other type of particles are part of
the static cluster partially and partially they are moving randomly (as shown in black arrow in Fig. \ref{fig:fig1}(e))
. In the bottom panel we show 
the zoomed version of the same snapshot for small part of the total system. Which shows that
orientation of particles inside the cluster is random. 
In Fig. \ref{fig:fig4} we plot
$P(v)$, which shows a broad distribution for type one  particles and 
two distinct peaks at $v=0.5$ and $v=0.0$ for type two  particles. Which again due to
 static clusters. Size of the peak at $v=0.5$ is large for type one
particle in comparison to second type. Hence large number 
of type one particles are moving randomly. 
To  further characterise this phase also plot the four $P_{ij}(n)$'s.
 In this case the four
distribution are very different from the previous cases. We
find that $P_{11}(n)$ is Gaussian and shows a peak at some finite
value of $n$. That typically represent the mean number 
of particles in the interaction radius. This is due to
presence of type one particle in the static regions of 
cluster formed by second type particles, which acts like nucleation site for particle of type one. 
The $P_{12}(n)$,  shows a broad distribution which confirms 
that the cluster of particle of type one has another particle too (due to fixed second type particle). The 
 two other distributions $P_{22}(n)$ and $P_{21}(n)$ are similar. When plotted on 
log-log scale, the two $P_{21}$ and $P_{22}$
shows  exponential tail with $n_0 \approx 5$

\subsection{Dynamics of particle in DM and DPS phase}
\begin{figure}[ht]
\centering
\includegraphics[width=0.8\linewidth]{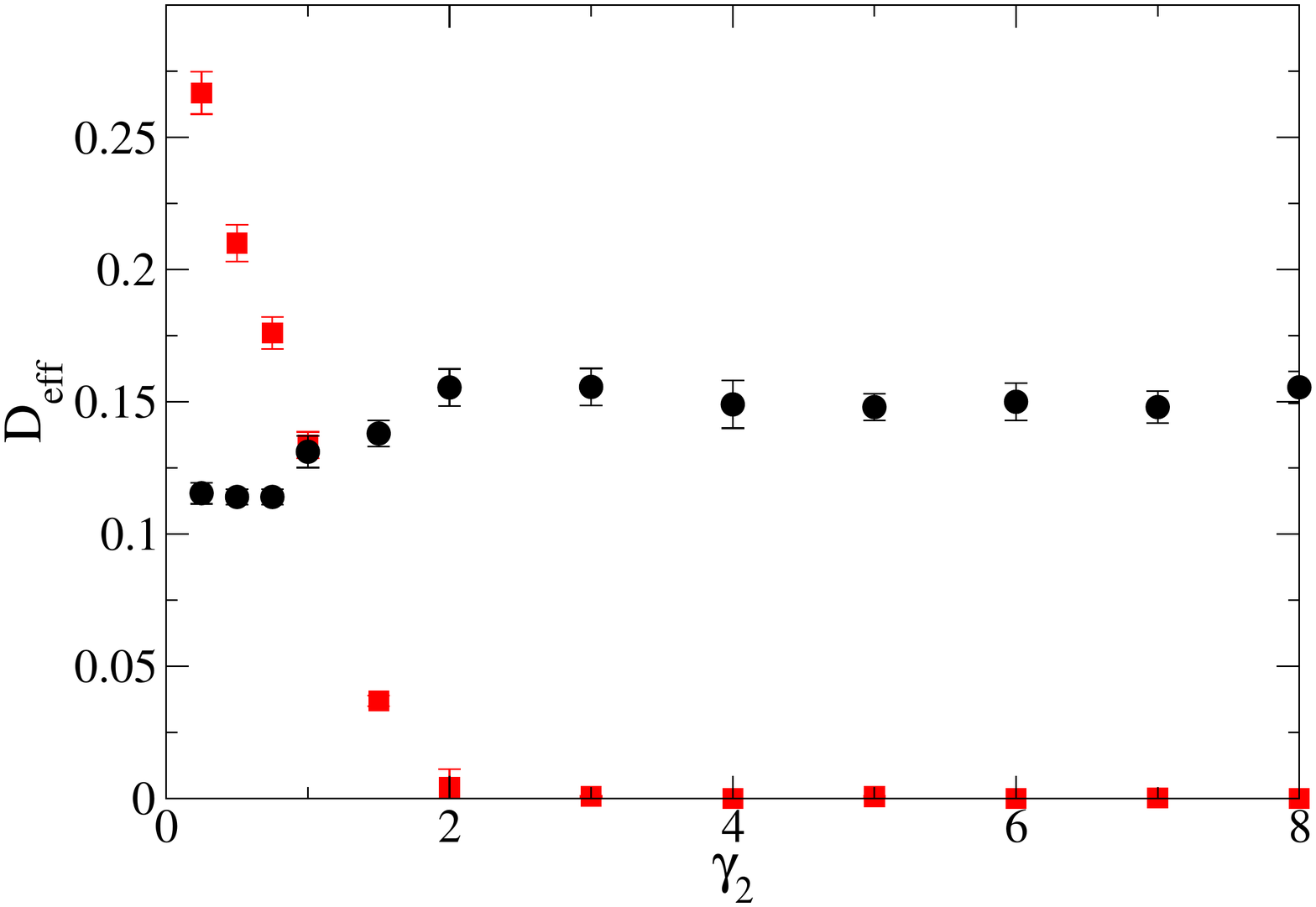}6
	\caption{Plot of effective diffusivity $D_{eff}$ vs. $\gamma_2$ in the disordered region $\eta=0.65$ for two types 
	of particles. $\circ$, $\square$ is for particle of type one and two respectively. Other parameters are same 
	as given in Fig. \ref{fig:fig1}}
\label{fig:fig7}
\end{figure}

We also characterise  the dynamics
of both types of particles in DM and DPS phase. We first calculate the 
mean square displacement MSD $\Delta_{i}(t) = <|{\bf r}_{i}(t+t_0)-{\bf r}_{i}(t)|^2>$, where 
$i=1,2$ for particle of type one and two respectively. $<.>$ is over all the particles of same type and many reference
time $t_0$.  MSD is 
 calculated  for $\eta=0.62$  and for different $\gamma_2$. 
In the disordered region or when $\eta>0.6$ 
we find that for both $\Delta_{i} (t) \simeq t$, 
which suggest the diffusive 
behavior of  particles. We further  estimate the effective diffusion coefficient 
$D_{eff} = \lim_{t \rightarrow \infty} \frac{\Delta (t)}{4 t}$.
Hence in Fig. \ref{fig:fig7} we plot the effective diffusion coefficient $D_{eff}$ vs.
$\gamma_2$.  For small $\gamma_2$ diffusivity of both 
types particle is finite but as we increase $\gamma_2$, diffusivity of
second type particle is almost zero. Which suggest static clusters of second
type particle as found in DPS phase. 

\section{Discussion\label{Discussion}}

We have studied the binary mixture of polar self-propelled particles with variable
speed. Speed of the particle depends on its neighbors and its maximum in well 
aligned region and almost zero in random disorder region. Dependence of local
speed on local orientation is controlled by a variable speed parameter $\gamma$. The model
is motivated with experiments on fish school where speed of individual fish 
depends on their neighbors. We mix the two different types of particles 
with two different $\gamma$ values. One of the $\gamma_1$ is fixed to $1$ and another 
$\gamma_2$ is varied from $0$ to $8$. For $\gamma=0$ model reduces to constant speed model.
Steady state behavior of the system is studied for different combination of $(\gamma_2, \eta)$.
For all set of $\gamma$'s system shows a transition from disordered state to
ordered state. We find four different phases: (i) ordered phase separated (OPS) when noise 
is small and difference in two $\gamma$'s is large. In this phase starting from 
random mixed phase both types of particle phase separate and moves in different clusters. 
(ii) ordered mixed phase (OM), when the difference is $\gamma$ is small then all 
the particles moves in well ordered cluster but in a single cluster both types of 
particles are present, (iii) disorder mixed phase (DMP), 
when $\eta$ is large and different in two $\gamma$ is small then both orientation and density
order parameter is small and both types of particles remain is mixed phase and have random orientation.
(iv) disorder phase segregated (DPS), when one of the $\gamma$ is large and noise in also 
large then local orientation is small hence for larger $\gamma_2$, speed of the second 
type of particle is almost zero and hence they form static clusters and speed of the first types 
particle varies linearly with local polarisation. Second type of particle which form static cluster
with completely random orientation acts like nucleation site and then first type of particle
come in contact with the static cluster they also form 
small clusters there. Which leads to a characteristic cluster size for first type 
particle. 
Hence our study shows appearance of different phases in binary active mixture 
of SPP's with variable speed. The variable speed parameter introduced here
can be thought of as characteristic of particle. Hence our study give insight 
to phase separation in different particle types. Also it opens new direction 
to study these systems in detail. It is also interesting to study 
the ordering kinetics \cite{ajbray} of two
types particle  in such mixture.

\section{Mean-field order-disorder transition using coarse-grained hydrodynamics\label{Mean-field order-disorder transition using coarse-grained hydrodynamics}}
We begin by defining the coarse-grained local density field  for two types of particles.  For particle of
type one
\begin{equation}
	\rho_1({\bf r},t)=\sum_{i=1}^{N_1}\delta({\bf r}-{\bf r}_i)
	\label{eq5}
\end{equation}
and for type two
\begin{equation}
	\rho_2({\bf r},t)=\sum_{j=1}^{N_2}\delta({\bf r}-{\bf r}_j)
	\label{eq6}
\end{equation}
where ${\bf r}_i$ and ${\bf r}_j$ is position vector of particle of type one and two.  
Similarly we define the local coarse-grained polarization  field as 
\begin{equation}
{\bf P}({\bf r}, t)={\frac{{\sum}^N_{k=1}\bf n_i(t)\delta(\bf r-r_{k})}{\rho({\bf r},t)}},
	\label{eq7}
\end{equation}
where $N= N_1+N_2$ and $\rho=\rho_1+\rho_2$ and $\sum_k$ is over all the particles.
Using the update rules of position Eq. \ref{eqn1}, orientation Eq. \ref{eqn2}  and using  the same  analysis as in 
\cite{shradhanjop, shradhathesis, shradhapre2012, shradhamanna},  we now  write the 
stochastic partial differential equations of motion for  two coarse-grained densities 
$\rho_1({\bf r},t)$ and $\rho_1({\bf r},t)$ and polarisation vector $P({\bf r}, t)$ as defined in 
Eqs. \ref{eq5}, \ref{eq6}, \ref{eq7}. We begin with density equation for type one particle.

\begin{equation}
	\rho_1({\bf r},t+\Delta{t})-\rho_1({\bf r},t))  = {\sum}^{N_1}_{i=1}[\delta({\bf r}-{\bf r}_{i}(t+\Delta{t})]-\delta[{\bf r}-{\bf r}_{i}(t)] \notag
\end{equation}

\begin{equation}
	  \Rightarrow -{\sum}^{N_1}_{i=1}v_{i}(t)\bf n_{i}(t)\nabla\delta[{\bf r}-{\bf r}_{i}(t)]+\frac{1}{2}{\sum}^{N_1}_{i=1} {\sum}^{N_1}_{i'=1}v^2_{i}{\bf n_{i}}(t){\bf n_{i'}}(t):\nabla_i\nabla_i'\delta[{\bf r}-{\bf r_{i}}(t)] 
\label{eq8}	
\end{equation}

Now we substitute for variable speed from Eqs. \ref{eqn3} and \ref{eqn4}  and 
dividing by $\Delta{t}$ and taking the limit $\Delta{t}\longrightarrow{0}$
\begin{eqnarray}
	\partial_t \rho_1({\bf r}, t) & =  v_0 {\sum}^{N_1}_{i=1}[ {[\frac{{\sum}^{R_0}_{j} {\bf n}_j}{N_i}]}^{\gamma_{1}} v_{i}(t){\bf n}_{i}(t)\nabla\delta[{\bf r}-{\bf r}_{i}(t)]  \notag
		 & + v_0^2\frac{1}{2}{\sum}^{N_1}_{i=1} {\sum}^{N_1}_{i'=1}[\frac{{\sum}^{R_0}_{j} {\sum}^{R_0}_{j'} {\bf n}_j {\bf n}_j'}{N_i N_i'}]^{\gamma_{1}} :\nabla_i\nabla_i'\delta[{\bf r}-{\bf r}_{i}(t)] 
\label{eq9}
\end{eqnarray}

Here the operator ``$:$'' is the double dot (or colon) product defined by 
${\bf ab:cd} =\sum_\alpha\sum_\beta{a^{\alpha}b^{\beta}c^{\alpha}d^{\beta}}$, with indexes $\alpha$ and $\beta$ 
indicating the vector components $(1,2)$. 
The expansion in equation Eq. \ref{eq8} and \ref{eq9} 
is valid for small values of the order parameter field and small particle speed, 
such that the displacement per time step is much smaller than the interaction range.
We now use  mean-field approximation and replace the summation over particle inside the 
interaction radius $R_0$ by mean value of polarisation we find the final equation for
the density field $\rho_1$ as 
\begin{equation}
	{\partial_{t}\rho_1}=-\nabla\cdot(v_1 {\bf P} \rho_1)+ D {\nabla}^2(v_1^2 \rho_1)
	\label{eq10}
\end{equation}
where $v_1$ is self-propulsion speed of particle and in general depends on local polarisation.
In the same manner we can also derive the density equation for second type particle 
\begin{equation}
	{\partial_{t}\rho_2}=-\nabla\cdot(v_2 {\bf P} \rho_2)+ D {\nabla}^2(v_2^2 \rho_2)
	\label{eq11}
\end{equation}
Where $D$ is introduced as diffusion coefficient, which is function of microscopic parameters. Here we assume 
it to be constant.
Now we come to the equation for the polarisation order parameter. 
 After a long but straight-forward calculation as in \cite{shradhanjop, shradhathesis, shradhapre2012, shradhamanna, tonertu} and in 
 the mean-field limit as for density equations \ref{eq10} and \ref{eq11}, 
 the polarisation equation will have mainly following terms

\begin{equation}
\partial_t{\bf P}=(\alpha_1(\rho, \eta)-\alpha_2 {\bf P}\cdot{\bf P}){{\bf P}} -\frac{v_1}{2}\nabla\rho_1 -\frac{v_2}{2}\nabla\rho_2 +\lambda({\bf P}\cdot\nabla){\bf P}+k\nabla^2{\bf P} + {\bf H}
\label{eq12}
\end{equation}
where on the R. H. S.  of above equation the first term  is the polynomial term which determines the 
order-disorder mean field transition. The second and third terms, the two gradients 
in density, is the change in local polarisation due to the variation in density of two types of particles, $\lambda$ term 
in the non-linear term and coefficient $\lambda$ in general depends on the microscopic parameters {\it viz} (mean density 
$\rho_0$, speed ${\bf v_0}$  etc.). In general we have three kinds of non-linearities as given 
in  \cite{shradhanjop, shradhathesis, shradhapre2012, shradhamanna, tonertu} but we keep only 
one of the relevant one as shown in recent study of \cite{tonertupre2018}. The last term is the diffusion 
of local polarisation  and $\bf H$ is the stochastic noise term with 
 $H=\sqrt{\rho}{\bf M \bf h}$
 Here $\bf h(r,t)$ is a vector field of unit length and random orientation, delta correlated in space and time, while $\bf M(r,t)$ is a $2\times{2}$ tensor field satisfying $\bf M^2=1$.
Here we are mainly interested in the mean-field order-disorder transition which is mainly
predicted by $\alpha_1(\rho, \eta)$.When derived from microscopic and and in the mean-field limit  $\alpha_1(\rho, \eta) = (\rho_0-4 \eta^2)$.  
Hence we find that $\alpha_1$ is only function of mean density $\rho_0$ and noise $\eta$ 
and is independent of the variable speed parameter $\gamma$.
Which confirms the disorder-to-order transition remains 
invariant with respect to the $\gamma$ as found in our numerical simulation.
\section{Linearised study of hydrodynamic equations of motion\label{Linearised study of hydrodynamic equations of motion}}
In this section we will do the linearised study of hydrodynamic equations of motion 
derived for the two density fields and polarisation Eqs. \ref{eq10}, \ref{eq11} and \ref{eq12}.
We take the mean-field approximation so that the speed of two particles can be replaced by $v_{1,2} = v_0 (p_0)^{\gamma_1, \gamma_2}$, where $p_0 = \sqrt{\frac{\alpha_1}{\alpha_2}}$ in Eq. \ref{eq12}.
Using the two density equation we write the equation in terms of difference in the density
of both types of particle $\Delta \rho = \rho_1-\rho_2$
\begin{equation}
\partial_t\Delta \rho=-v_1\nabla\cdot({\bf P}\rho_1)+v_2\nabla\cdot({\bf P}\rho_2)+D\nabla^2 \Delta \rho
\label{eq13}
\end{equation}
which is further equal to
\begin{equation}
\partial_{t}\bigtriangleup\rho=-\bigtriangleup{v}\nabla\cdot({\bf P}\rho_1)-v_2\nabla\cdot({\bf P}\bigtriangleup\rho)+D\nabla^2\bigtriangleup\rho
\label{eq14}
\end{equation}  
where $ \Delta v = v_1-v_2$. 
and equations for the density of particle of type one is same as in Eq. \ref{eq11}. In the same manner we write the
polarisation equation also in terms of $\Delta \rho$ and $\Delta v$
\begin{equation}
	\partial_t{\bf P}=(\alpha_1(\rho, \eta)-\alpha_2 {\bf P}\cdot{\bf P}){{\bf P}} -\frac{\overline v}{2}\nabla ( v_1 \rho_1) +\frac{v_1}{2}\nabla( \Delta \rho) - \frac{\Delta v}{2} \nabla (\Delta \rho)  +\lambda({\bf P}\cdot\nabla){\bf P}+k\nabla^2{\bf P} + {\bf H}
	\label{eq15}
\end{equation}

The homogeneous steady state solution of above three equations for $\Delta \rho$, $\rho_1$ and ${\bf P}$ is
$\Delta \rho=0$, $\rho_1=0$ and ${\bf P}  = \sqrt{\frac{\alpha_1}{\alpha_2}}{\widehat{\bf x}}$ (the direction of broken symmetry  along  $x-$axis). We add small perturbation about the 
above homogeneous solution hence  
$\Delta \rho_1=\delta \Delta \rho$, 
$\rho_1=\rho_{10} + \delta\rho_1$
 and 
${{\bf P}}=(p_0+\delta{p_\parallel}){\widehat{\bf x}}+(\delta{p}_{\perp}){\widehat{\bf y}}$, where $p_0= \sqrt{\frac{\alpha_1}{\alpha_2}}$ and 
$\delta \Delta \rho =\delta  \rho_1 - \delta \rho_2$  is the fluctuation is the two densities about their mean values.
Now we write the equations for small perturbations in four fields $\Delta \rho$, $\delta \rho_1$, $\delta p_{\parallel}$ and $\delta p_{\perp}$.
We first write the equation for $p_{\parallel}$ first
\begin{equation}
\partial_t\delta{p}_\parallel=(\alpha_1(\rho_0)-\alpha_2(p
_0^2+2p_0\delta{p}_\parallel))(p_0+\delta{p}_\parallel)-\frac{v_1}{2}\nabla\delta\rho_1-\frac{v_2}{2}\nabla(\rho_1-\bigtriangleup\rho)+\lambda(p_0\delta_x)\delta{p}_\parallel+k\nabla^2\delta{p}_\parallel
\label{eq16}
\end{equation}
If we  ignore the higher order gradients terms 
we find in the steady state
\begin{equation}
\delta{p}_\parallel=-\frac{v_1}{4\alpha_1}(v_1+v_2)\nabla_x\delta\rho_1+\frac{v_2}{4\alpha_1}\nabla_x(\bigtriangleup\rho)\\
\label{eq17}
\end{equation}
Now we write the equations for the small fluctuations in other three field and substitute the expression for $\delta P_{\parallel}$ 
from Eq. \ref{eq16}
\begin{equation}
\partial_t\delta{p}_\perp=-\frac{v_1}{2}\delta_y\delta\rho_1-\frac{v_2}{2}\delta_y(\delta\rho_1-\bigtriangleup\rho)+\lambda(p_0\delta_x)\delta{p}_\perp+k\nabla^2\delta{p}_\perp
\label{eq18}
\end{equation}
substitute from Eqs. \ref{eq9}, \ref{eq10} and \ref{eq13},
from equation \ref{eq17}, we solve for $\partial{p}_{\parallel}$ and substitute in equation \ref{eq18}
\begin{equation}
\partial_t\delta {p}_\perp=-v_1p_0\delta_x\delta\rho_1+\frac{v_1\rho_
{10}}{4\alpha_1}\overline{v}\delta_x^2\delta\rho_1-\frac{v_1v_2\rho_10}{4\alpha_1}\delta_x^2\bigtriangleup\rho-v_1\rho_{10}\delta_y\delta{p}_\perp+D\nabla^2\delta\rho_1\\
\label{eq19}
\end{equation}
where $\overline{v} = v_1+v_2$, 
similarly we write equations for $\partial_{t}\delta\rho_{1}$ and $\partial_{t}\Delta\rho$
\begin{equation}
\partial_t\delta\rho_1=-v_1\delta_x((p_0+\delta{p}_\parallel)(\rho_{10}+\delta \rho_1))-v_1\delta_y(\delta{p}_\perp\rho_{10})+D\nabla^2\delta\rho_1\\
\label{eq20}
\end{equation}
\begin{equation}
\partial_t\bigtriangleup\rho=-\bigtriangleup{p_0}\delta_x\delta\rho_1+\frac{\bigtriangleup\overline{v}\rho_{10}\delta_x^2\delta\rho_1\overline{v}}{4\alpha_1}-\frac{\bigtriangleup{v}\rho_{10}v_2\delta_x^2\bigtriangleup\rho}{4\alpha_1}-\bigtriangleup\rho_{10}\delta_y\delta{p}_\perp-v_2p_0\delta_x\bigtriangleup\rho+D\nabla\bigtriangleup\rho\\
\label{eq21}
\end{equation}
now taking the Fourier transformation equation \ref{eq18} \ref{eq19} and \ref{eq20} by using 
\begin{equation}
Y=\begin{bmatrix}
\partial\rho_1\\
\partial{p_{\perp}}\\
\Delta\rho
\end{bmatrix},
 Y(k,S)=\int{Y(r,t)}\exp{(S{t}-i\overline{k}\cdot\overline{r})} d\overline{r}dt
 \label{eq22}
\end{equation} 
and write in Fourier space
\begin{equation}
(S+v_1+p_0iq_x+\frac{v_1\rho_{10}\overline{v}q_x^2}{4\alpha_1}+Dq^2)\delta\rho_1-(\frac{v_1v_2\rho_{10}q_x^2}{4\alpha_1})\bigtriangleup\rho+v_1\rho_{10}iq_y\delta{p}_\perp=0\\
\label{eq23}
\end{equation}
\begin{equation}
(\bigtriangleup{v}p_0iq_x+\frac{\bigtriangleup{v}\overline{v}\rho_{10}q_x^2}{4\alpha_1})\delta\rho_1+(S-\frac{\bigtriangleup{v}v_2\rho_{10}q_x^2}{4\alpha}+Dq^2+v_2p_0iq_x)\bigtriangleup\rho+\bigtriangleup{v}\rho_{10}iq_y\delta{p}_\perp=0 \\
\label{eq24}
\end{equation}
\begin{equation}
(\frac{v_1iq_y}{2}+\frac{v_2iq_y}{2})\delta\rho_1-(\frac{v_2iq_x}{2})\bigtriangleup\rho+(S-\lambda{p_0}iq_x+kq^2)\bigtriangleup{p}_\perp=0\\
\end{equation}
$M\times{Y}=0$ where
\begin{equation}
M=\begin{bmatrix}
(S+v_1p_0iq_x+\frac{v_1\rho{10}\overline{v}q_x^2}{4\alpha_1}+Dq^2)&(-v_1v_2\rho_{10}q_x^2/4\alpha_1)&(v_1\rho_{10}iq_y)\\
(\bigtriangleup{v}p_0iq_x+\frac{\bigtriangleup\overline{v}\rho_{10}q_x^2}{4\alpha_1})&(S-\frac{\bigtriangleup{v}v_2\rho{10}q_x^2}{4\alpha_1}+Dq^2+v_2p_0iq_x)&(\bigtriangleup\rho_{10}iq_y)\\
(\frac{\overline{v}}{2}iq_y)&(-\frac{v_2}{2}iq_y)&(S-\lambda{p_0iq_x}+kq^2)
\end{bmatrix}
\begin{bmatrix}
\delta\rho_1\\
\delta{p}_\perp\\
\bigtriangleup\rho
\end{bmatrix}
\label{eq25}
\end{equation}
Now we focus along the ordering direction $q_y=0 , \theta=0, q_x=q$ and $det[M]=0$\\
Eq \ref{eq25} can be solved for modes by det$[M]=0$
\begin{equation}
0=\begin{bmatrix}
(S+v_1p_0iq+\frac{v_1\rho{10}\overline{v}q^2}{4\alpha_1}+Dq^2)&(-v_1v_2\rho_{10}q^2/4\alpha_1)&0\\
(\bigtriangleup{v}p_0iq+\frac{\bigtriangleup\overline{v}\rho_{10}q^2}{4\alpha_1})&(S-\frac{\bigtriangleup{v}v_2\rho{10}q^2}{4\alpha_1}+Dq^2+v_2p_0iq)&0\\
0&0&(S-\lambda{p_0iq}+kq^2)
\end{bmatrix}
\label{26}
\end{equation}
One of the mode
$\Rightarrow{S=\lambda{p}_0iq-kq^2}$ Damped-diffusive oscillatory modes and other two modes are given by
\begin{equation}
0=\begin{bmatrix}
(S+v_1p_0iq+q^2(D+\frac{v_1\rho_{10}\overline{v}}{4\alpha_1})&(-v_1v_2\rho_{10}q^2/4\alpha_1)\\
(\bigtriangleup{v}p_0iq+\frac{\bigtriangleup{v}\overline{v}\rho_{10}q^2}{4\alpha_1})&(S+v_2p_0iq+q^2(D-\frac{v_2\rho_{10}\bigtriangleup{v}}{4\alpha_1})
\end{bmatrix}
\label{eq27}
\end{equation}
lets define  $D_{+}=q^2(D+\frac{v_1\rho_{10}\overline{v}}{4\alpha_1})$ and $D_{-}=q^2(D-\frac{v_2\rho_{10}\bigtriangleup{v}}{4\alpha_1})$
Hence we have two solutions for  $S$, $\mathcal{R}e (S)>0$ (mode is unstable) and when $\mathcal{R}e (S)<0$ (then it is stable)
\begin{equation}
(S+v_1p_0iq+q^2D_+)(S+v_2p_0iq+q^2D_-)+\frac{v_1v_2\rho_{10}\bigtriangleup{v}p_0iq^3}{4\alpha_1}+\frac{\bigtriangleup{v}\overline{v}v_1v_2\rho_{10}^2q^4}{(4\alpha_1)^2}=0\\
\label{eq28}
\end{equation}
\begin{equation}
{2S}_{\pm}=
[(v_1p_0iq+D_{+q}^2)+(v_2p_0iq+D_-q^2)]\pm[(v_1p_0iq+D_{+}q^2)-(v_2p_0iq+D_-q^2)]
\label{eq29}
\end{equation}
\begin{equation}
{S}_{+}=
[(v_1p_0iq+D_{+q}^2)+(v_2p_0iq+D_-q^2)]+[(v_1p_0iq+D_{+}q^2)-(v_2p_0iq+D_-q^2)]
\label{eq29}
\end{equation}
The $-Ve$ root $S_-=-2(v_1p_0iq+D_+q^2)\rightarrow$, is always stable
The $+Ve$ root $S_+$ can become unstable since $p_0<1$ as if
$\bigtriangleup{v>\frac{4D\alpha_1}{v_2\rho_{10}}}\Rightarrow(\frac{•8D\alpha_1}{v_2\rho_0})$. Hence criticality 
arise at
$\bigtriangleup{v}=\frac{8D\alpha_1}{\rho_0v_0p_0^{\gamma_2}}$
As $\gamma_2$ increases the difference in two speeds $\Delta v$ also increases which leads to more
instability. Similar trend is obtained in our numerical simulation when difference in two $\gamma{'s}$ large than order-homogeneous state is unstable.







\end{document}